\documentstyle[aps,prl,twocolumn,epsfig,floats]{revtex}
\newcommand{\be}{\begin{eqnarray}}
\newcommand{\ee}{\end{eqnarray}}

\begin{document}
\draft
\wideabs{
\title{Thouless Energy and Correlations of QCD Dirac Eigenvalues}
\author{J.C. Osborn and J.J.M. Verbaarschot}
\address{%
Department of Physics and Astronomy, SUNY, 
Stony Brook, New York 11794}
\date{\today}
\maketitle
\begin{abstract}
Eigenvalues and eigenfunctions of the QCD Dirac operator are studied for
an instanton liquid partition function. We find that
for energy differences $\delta E$ 
below an energy scale $E_c$, identified as the Thouless energy, the eigenvalue 
correlations are given by Random Matrix Theory.
The value of $E_c$ shows a weak volume dependence for 
eigenvalues near zero and is consistent with a scaling
of $E_c \sim 1/L^2$ in the bulk of the spectrum in 
agreement with estimates from chiral perturbation theory, that 
$E_c/\Delta \approx F_\pi^2 L^2/\pi$ (with average level spacing $\Delta$). 
For $\delta E> E_c$ the number variance 
shows a linear dependence. For the wave functions we find a small nonzero 
multifractality index. 
\end{abstract}
\pacs{PACS numbers: 12.38.Lg, 71.30+h, 72.15.Rn, 11.30.Rd,  05.45.+b, 12.38.Gc}
}

\narrowtext

Random matrix theories have been applied to many aspects of mesoscopic systems
(see \cite{HDgang,Beenreview,Montambaux} for recent reviews). In particular,
eigenvalue correlations have received a great deal of attention in
this context. Two important
energy scales have been identified: i) the Thouless energy, defined as the 
inverse diffusion time of an electron 
through the sample, i.e.
\be
E_c = \frac {\hbar D}{L^2},
\ee
where $D$ is the diffusion constant, and ii) the energy $\hbar/\tau_e$, 
where $\tau_e$ is the elastic collision time. 
Eigenvalue correlations on a scale $\delta E$ 
can then be classified according to following three
regimes: the ergodic regime for $\delta E < E_c$, the diffusive or 
Altshuler-Shklovskii \cite{Altshuler} 
regime for $E_c < \delta E < \hbar/\tau_e$ and the
ballistic regime for $\delta E > \hbar/\tau_e$
(see \cite{Altland} for recent work on this topic.).

The eigenvalue correlations can be measured conveniently
by the number variance, $\Sigma_2(n)$. This statistic is defined
as the variance of the number of levels in an interval that contains $n$
levels on average. In the ergodic regime, eigenvalue correlations
are given by the invariant random matrix ensembles with $\Sigma_2(n) \sim 
(2/\beta \pi^2) \log(n)$. In the diffusive regime the situation is
more complicated. For a critical system, with a localization length that 
scales with the size of the sample, it is argued \cite{Altshuler}
that $\Sigma_2(n) = \chi n$ (with $\chi < 1$), whereas for weaker disorder
the expectation is that $\Sigma_2(n) \sim n^{d/2}$. For a critical system, the
slope of the number variance has been related to the multifractality index of
the wavefunctions \cite{Chalker-kravtsov}. 

In this letter we wish to investigate to what extent 
such scenarios are realized in QCD.
We will investigate eigenvalues of the Euclidean Dirac operator. Because of
the $U_A(1)$ symmetry they occur in pairs $\pm \lambda_k$ or are zero. 
What is of main interest are the eigenvalues near zero
which, for broken chiral
symmetry, are spaced as $\Delta= \pi/\Sigma V$ 
(the space time volume is denoted by $V$). 
This is based on the Banks-Casher formula
\cite{BC} according to which the order parameter of the chiral phase transition,
$\Sigma$, and the spectral density near zero are related 
by $\Sigma = \pi\rho(0)/V$. 
It is therefore natural to define the microscopic
limit, $V\rightarrow \infty$ with $u=\lambda V\Sigma $ kept fixed and
the associated microscopic spectral density \cite{SV}
\be
\rho_S(u) = \lim_{V\rightarrow \infty} \frac 1{V\Sigma} \langle
\rho(\frac u{V\Sigma})\rangle.
\label{rhosu}
\ee
There is ample evidence from lattice QCD \cite{Tilo,Ma}
and instanton liquid \cite{Vinst} simulations that $\rho_S(u)$
and other correlators on the scale of individual level spacings 
\cite{Halasz} are given by chiral Random Matrix Theory (chRMT), i.e.
RMT's with the chiral symmetries of the QCD partition
function. However, at scales beyond a few eigenvalue spacings
in both instanton \cite{Vinst} and lattice
QCD \cite{Tilo,Ma} simulations  the Dirac eigenvalues near zero show
stronger fluctuations than in chRMT.
This indicates the presence of an energy scale in QCD which may be
identified as the Thouless energy.
The interpretation of spontaneous chiral symmetry breaking as a delocalization
transition was  made earlier in \cite{shuryak}. By analogy with the
Kubo formula, $\Sigma$ plays the role of the conductivity \cite{shuryak}.
      
Because of the chiral symmetry of the Dirac operator and its spontaneous 
breaking, the eigenvalue correlations near zero in the ergodic domain 
are given by the  chiral ensembles \cite{SV,V}. This is the domain where
pion loops can be ignored. Its boundary is thus given by \cite{GL,LS}
a valence quark mass scale $m_c$ where the mass of 
the associated Goldstone 
boson, $\sqrt{m_c B}$ (according to the PCAC relation
$B = \Sigma/F_\pi^2$, where $F_\pi$ is the pion decay constant),
is of the order of the inverse linear dimension
of the box. This relation can be rewritten as \cite{vPLB}
\be
m_c = \frac 1{B L^2}.
\label{range}
\ee
It is therefore tempting to interpret $1/B$ as the diffusion constant.
Because $B$ is large on a hadronic scale ($B\approx 1660\, MeV$), the diffusion
constant is relatively small. With eigenvalue spacing 
$\Delta = \pi/\Sigma V$, this condition can
be rewritten in dimensionless form as
\be
g_c = \frac{m_c}\Delta = \frac{F_\pi^2}{\pi} L^2.
\label{gc}
\ee
Here, $g_c$ plays the role of the dimensionless conductivity. 
Another discussion of $F_\pi$ in terms of the 
conductivity is given in \cite{Stern}.
In lattice QCD, for an $Na^4$ lattice, 
this relation reads $g_c = F_\pi^2 a^2 \sqrt N/\pi$. On a
$16^4$ lattice with a lattice spacing of 0.2\,$fm$ this results in a
dimensionless Thouless energy of about one level spacing. The above 
discussion was for sea quark masses much less than the valence quark mass
scale $m_c$. Further subtleties arise in the quenched limit
(see \cite{osborn} for more details).   

The chiral Random Matrix Theories for $N_f$ massless quarks in the sector
of topological charge $\nu$ are defined by
the partition function \cite{SV,V}
\be
Z_{N_f,\nu}^\beta =
\int DW  {\det}^{N_f}\left (\begin{array}{cc} 0 & iW\\
iW^\dagger & 0 \end{array} \right )
e^{-{n \beta} {\rm Tr}V(W^\dagger W)},
\label{zrandom}
\ee
where $W$ is a $n\times (n+\nu)$ matrix.
The parameter $2n$
is identified as the dimensionless volume of space time. 
In this paper we only consider the chiral Gaussian Unitary Ensemble (chGUE)
with complex matrix elements $W$ ( $\beta = 2$) and a Gaussian 
potential $V(x) = \Sigma^2 x$. In this case $\rho_S(u)$
and all other correlation functions have been derived analytically 
\cite{V,VZ,Ma}. We only note that in the bulk of the spectrum the 
correlations are given by the invariant RMT's.
It can be shown \cite{Brezin-Zee,Damgaard} that, for $\beta =2$,
$\rho_S(u)$ and other correlators 
do not depend on the potential $V(x)$.
The application of chRMT to QCD has been 
put on a firm foundation by these and other universality proofs 
\cite{Dampart,Tilo-Guhr,Seneru}. Whether or not QCD is
in this universality class is a dynamical question that can only be proven
by explicit numerical simulations.

Our calculations will be performed using the instanton liquid model.
In this model the gauge field configurations are given by 
a superposition of instantons. 
The Euclidean QCD partition function is then approximated by
\be
Z_{\rm inst} = \int D\Omega \, {\det}^{N_f} (D + m) e^{-S_{\rm YM}},
\label{zinst}
\ee
where the integral is over the collective coordinates of the instantons
and the Dirac operator is denoted by $D$. For
each instanton we have $12$ collective coordinates (for three colors).
The Yang-Mills action is denoted by 
$S_{\rm YM}$.  The fermion determinant is evaluated in the space
of the fermionic zero modes of the instantons. 
In our calculations we use the standard 
instanton density $N/V = 1 $ (in units of $fm^{-1}$).
For further discussion of
this partition function, which obeys the flavor and 
chiral symmetries of the QCD
partition function, we refer to \cite{SS97}.

The partition function (\ref{zinst}) is evaluated by means of
a Metropolis algorithm. We perform on the order of 10,000
sweeps for each set of parameters. 
The eigenvalues and eigenvectors of the Dirac operator are calculated by
means of standard diagonalization procedures. In this letter we restrict 
ourselves to $N_f=0$. This is a natural choice because
our results are compared with ideas from 
disordered mesoscopic systems where no  
fermion determinant is present, and, moreover, it
allows us to study much larger volumes.

In order to separate the fluctuations of the eigenvalues from
the average spectral density, the spectrum is unfolded such that
the unfolded spectrum 
has unit average level spacing.
All our spectral observables are calculated with the unfolded spectrum.

\begin{figure}[!ht]
\centering
\includegraphics[width=70mm,height=50mm]{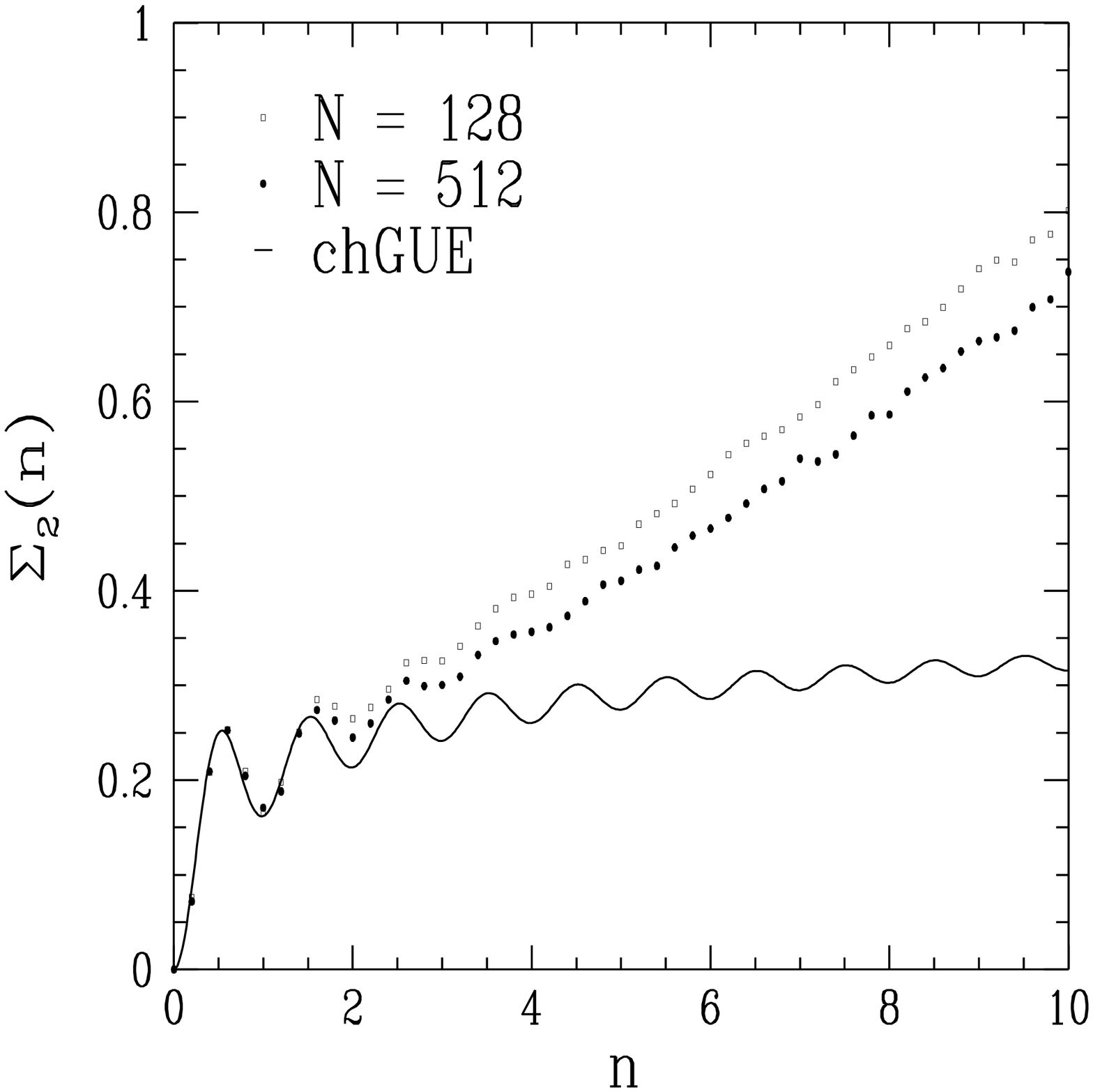}
\includegraphics[width=70mm,height=50mm]{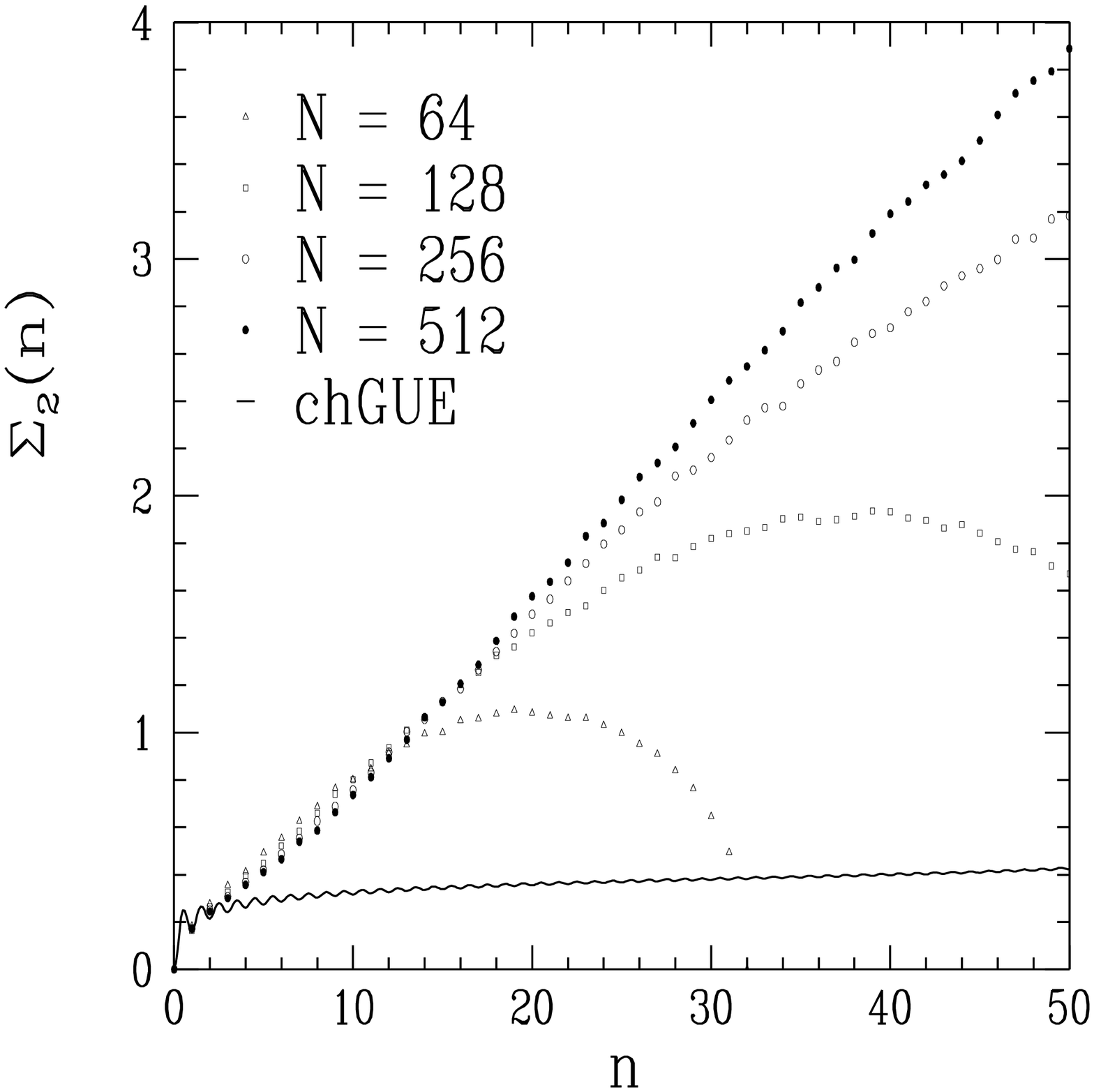}
\caption{The number variance $\Sigma_2(n)$ versus $n$ 
approximation for an interval starting at $\lambda = 0$. The total number
of instantons is denoted by $N$.}
\end{figure} 

\begin{figure}[!ht]
\centering
\includegraphics[width=70mm,height=50mm]{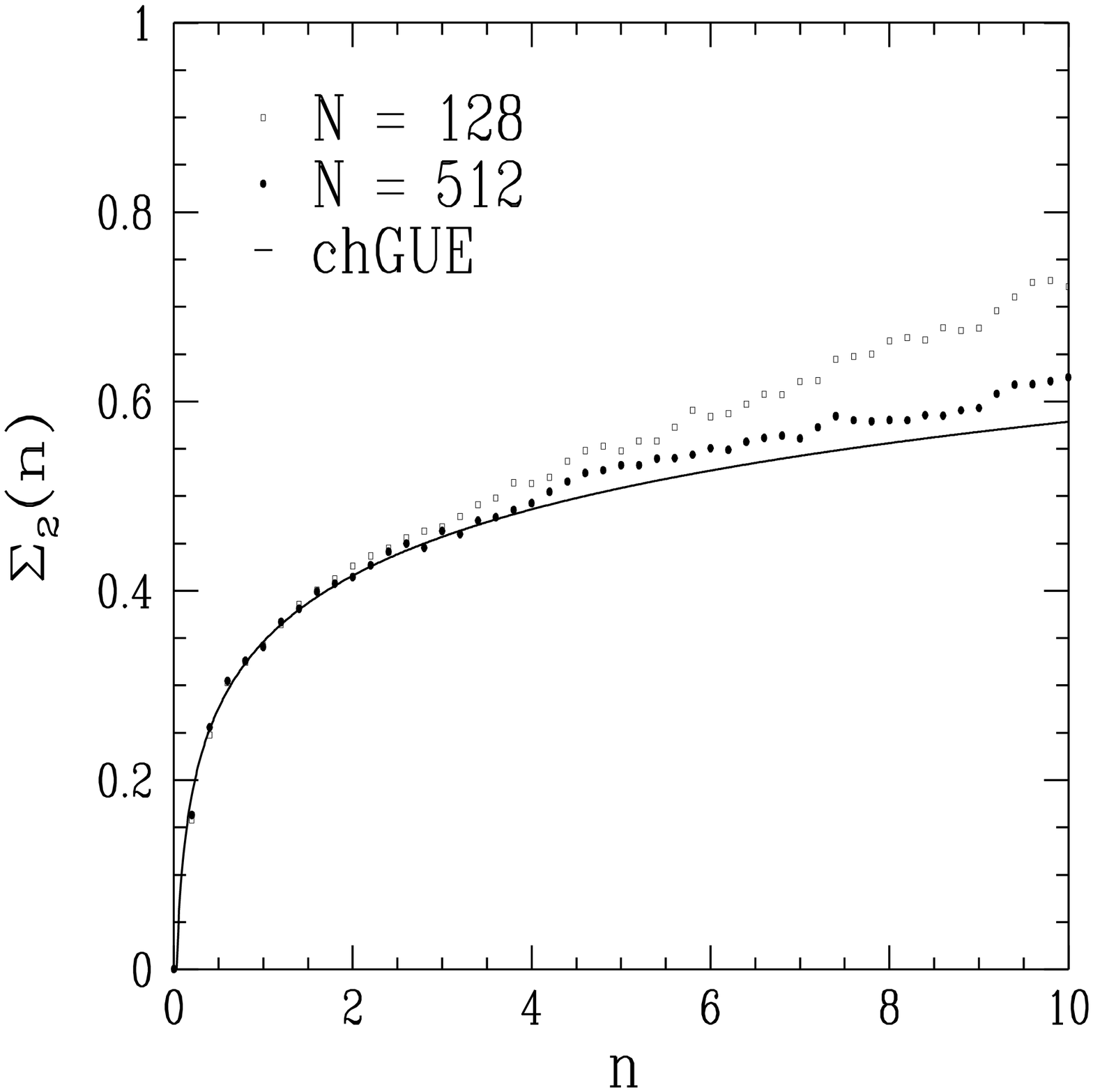}
\includegraphics[width=70mm,height=50mm]{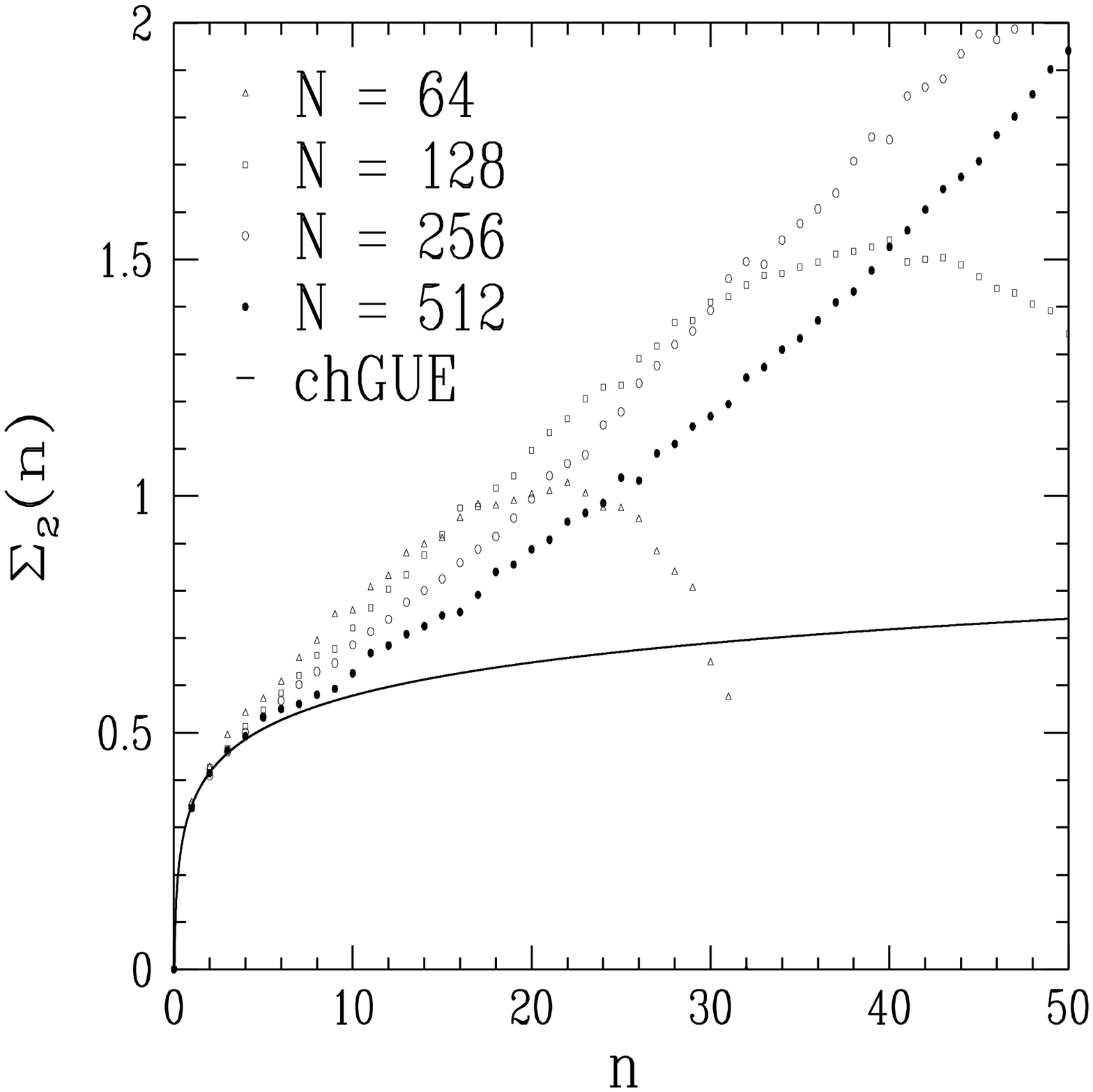}
\caption{The number variance $\Sigma_2(n)$ versus $n$ in the bulk of the
spectrum.}
\end{figure}

In Figs. 1 and 2 we show the number variance, $\Sigma_2(n)$ versus $n$ 
for various total numbers of instantons with instanton density $N/V= 1$. 
The chGUE  result for $\Sigma_2(n)$  \cite{Ma} is
depicted by the solid curve.
In both figures, the
upper figure (with only two volumes) 
is a blown-up version of the lower figure.
In Fig. 1, $\Sigma_2(n)$ is calculated for the interval 
starting at $\lambda=0$. Fig. 2 represents the 
number variance in the bulk of the spectrum obtained from an interval that
is symmetric about the average positive unfolded eigenvalue. In both cases
we observe a clear transition point $n_c$ below which the number variance
is given by RMT.
In Fig. 1,  the value of the crossover point, $n_c \approx 2$, 
depends only weakly on  the total number of instantons
(or the volume). This is not in agreement with
the theoretical expectation (\ref{gc}) 
that $n_c \approx F_\pi^2 \sqrt V/\pi$ in four dimensions. For 
$N/V = 1\, fm^{-4}$ the value of $n_c$ for $N$ instantons is given by
$n_c \approx 0.07 \sqrt N$ which is on the order of the results found in Fig. 
1. However in the bulk of the spectrum (Fig. 2), the value of $n_c$ is
consistent with a $\sqrt V$ scaling but the numerical constant appears 
to be larger than the above estimate.
This result is in agreement with the finding that 
correlations of lattice QCD eigenvalues 
are given by RMT up to distances of more than 100 spacings \cite{Halasz,Guhrp}.

Beyond the crossover point the number variance shows 
a linear behavior with a slope 
$\chi \approx 0.08$ for eigenvalues near zero and $\chi \approx 0.04$ 
in the bulk of the spectrum. 
The downward trend of the curves for larger values of $n$
is a well understood finite size effect. 
This prevents us from saying 
more about the ballistic regime, an energy scale of roughly 
the inverse distance between instantons. 

In the ergodic regime we expect that eigenvalue correlations are given
by the chiral random matrix ensembles. Indeed, 
both the microscopic
spectral density up to two level spacings and the nearest neighbor
spacing distribution are in perfect agreement with the chGUE.


The number of significant components of the wave function is  measured by the
participation ratio. Its inverse is defined as
\be
I_2(\lambda) = \langle \sum_k |\psi_k(\lambda)|^4 \rangle.
\ee 
The $\psi_k(\lambda)$ are the normalized $N$-component eigenfunctions
(corresponding to eigenvalue $\lambda$)
of the Dirac operator in the space of the fermionic zero modes of the 
individual instantons.
Results for $N I_2(\lambda)$ as a function of 
$\lambda$ are depicted in Fig. 3 (upper). 
The general impression is that the wave-functions are extended with 
an inverse participation ratio that is
not too different from the random matrix result (full line).
The eigenfunctions corresponding to small and large eigenvalues appear
to be somewhat more localized. A more definitive result for 
the character of the
wavefunctions follows from the scaling behavior of 
$I_2(\lambda)$ with the volume. 
A double logarithmic plot of 
$N I_2(\lambda)$ versus $N$ is  shown in Fig. 3 (lower). Results are given
both for the energy intervals $[0.1,0.2]$ (open circles) and $[0.7,0.8]$ (full 
circles). The first region corresponds to a part of the spectrum where
the number variance shows a linear behavior for the volumes shown in Fig. 1,
and the second region corresponds to the bulk of the spectrum.
\begin{figure}[!ht]
\centering
\includegraphics[width=70mm,height=50mm]{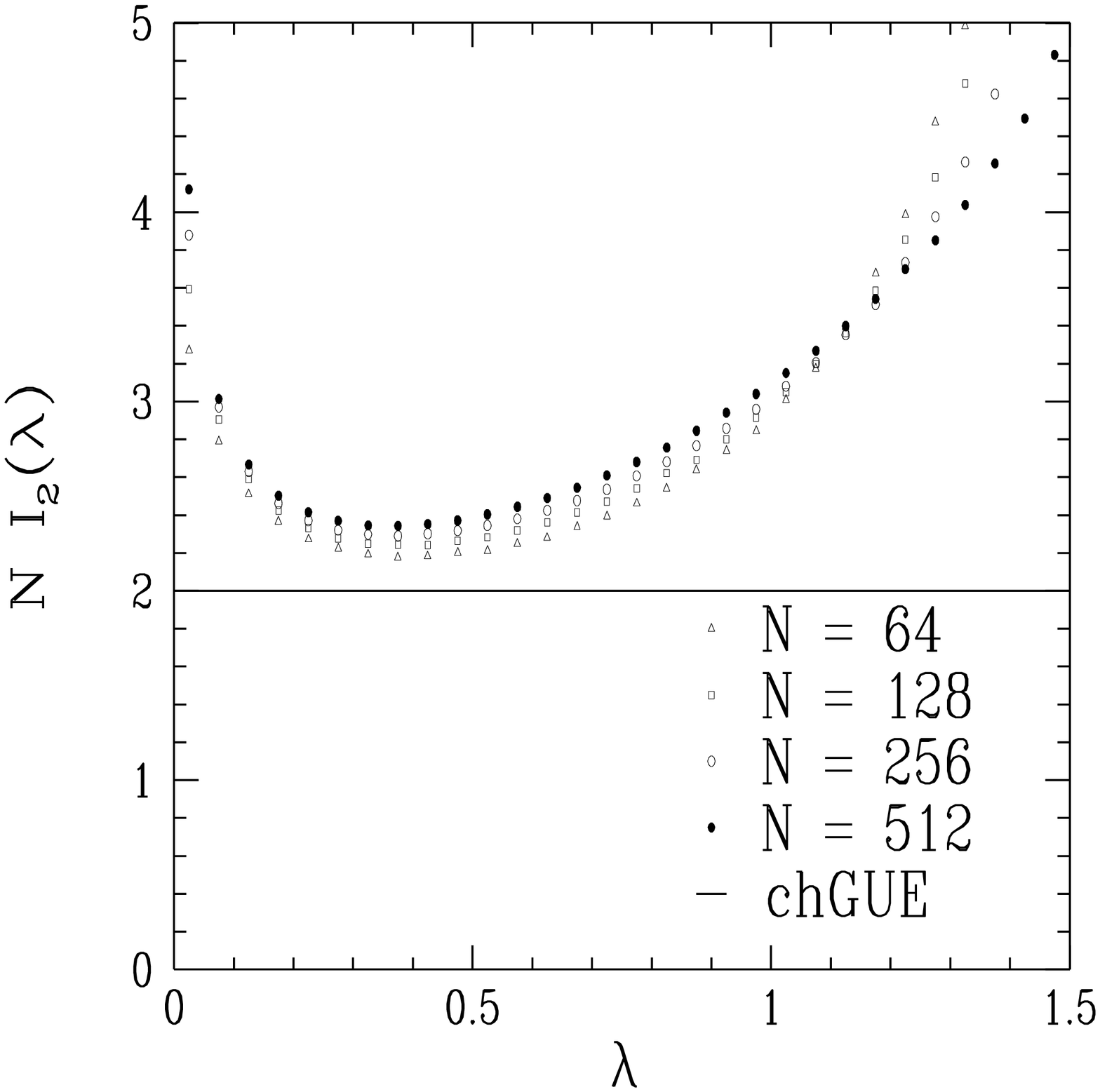}
\includegraphics[width=70mm,height=50mm]{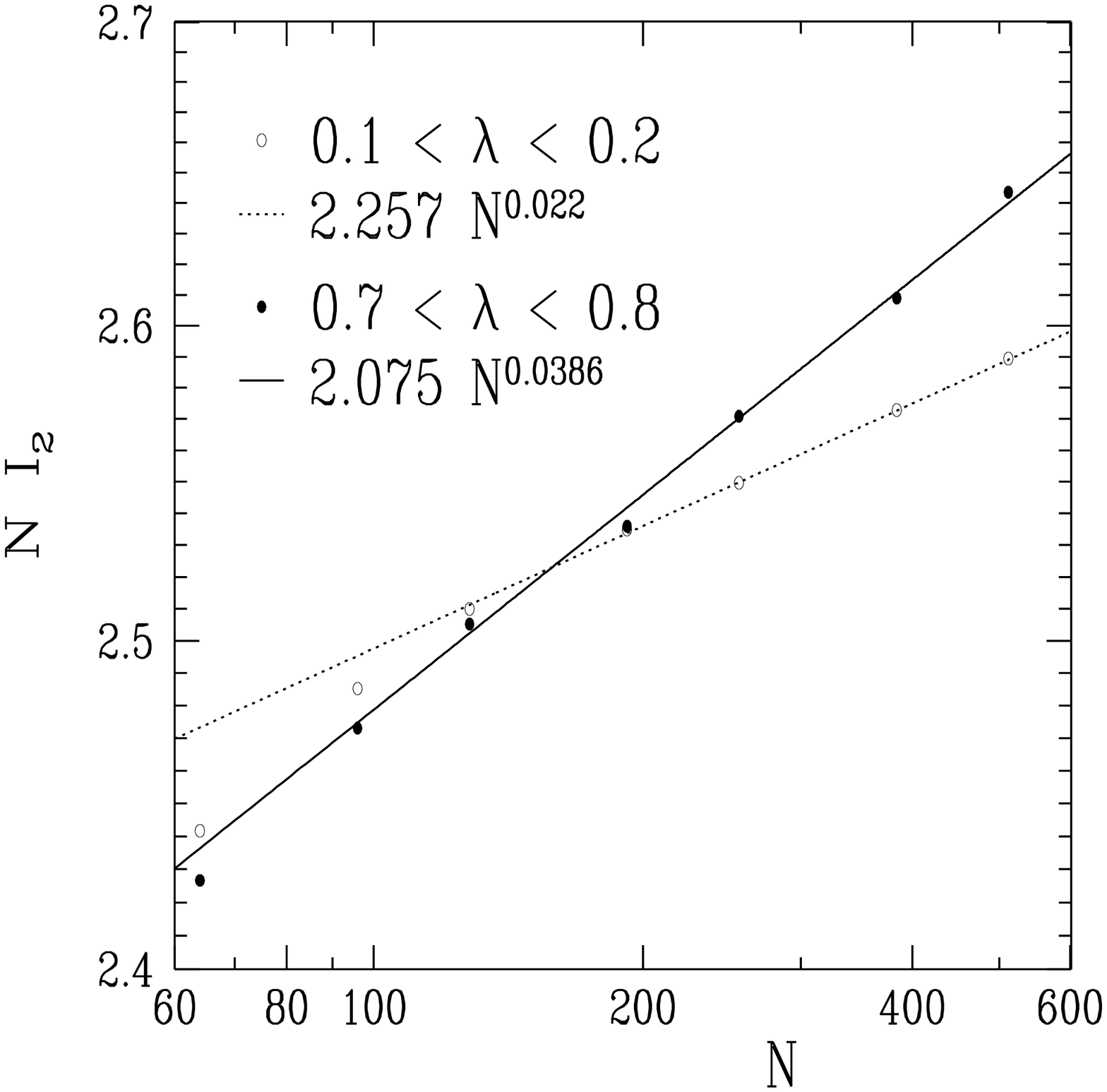}
\caption{The inverse participation ratio times $N$ 
versus the corresponding Dirac eigenvalues (upper),
and the scaling behavior versus $N$ (lower) 
for two energy intervals.
}
\end{figure}
The multifractality
index $\eta$ is defined by \cite{Chalker-kravtsov} 
\be
I_2 \sim V^{\eta/d-1}.
\ee
According to  \cite{Chalker-kravtsov} the 
value of $\eta=2\chi d$ (where $\chi$ is the slope of the 
linear piece in the number variance) in the critical domain. 
From the volume dependence of $I_2$ shown in 
Fig. 3 (lower) we find  values 
for $\eta/d$ of about 0.02 and 0.04 for the intervals
$[0.1,0.2]$ and $[0.7,0.8]$, respectively. These values are well below
the theoretical result of $2\chi$. Apparently, our ensemble of instantons
is not in the critical region. Our results should be contrasted with 
Wilson lattice QCD Dirac eigenfunctions which 
were found to be localized \cite{Janssen}.
We have no explanation for this discrepancy. 


In conclusion, we 
have identified an energy scale below which the eigenvalue correlations of
the QCD Dirac operator are given by chRMT. In
analogy with the theory of mesoscopic systems, this scale will be called
the Thouless energy. For eigenvalues near zero we find a Thouless energy
that only shows a weak volume dependence, whereas 
for eigenvalues in the bulk the Thouless energy 
scales roughly with the square root of
the volume in agreement with theoretical prediction.

For energy scales beyond the Thouless energy 
a linear $n$-dependence of the number variance is found. Our wave
functions show a small nonzero multifractality index which does not
obey the relation derived for critical mesoscopic systems.
Interesting connections with the scalar susceptibility and quenched chiral
perturbation theory will be discussed elsewhere \cite{Toublan}.

A. Smilga and Y. Fyodorov are thanked for stimulating discussions.
After the completion of this work we received a preprint (R. Janik et al.,
hep-ph/9803289) in which similar ideas were discussed. In that work the
diffusion constant is related to diffusion in a 4+1 dimensional space time.


\begin{references}

\bibitem{HDgang}T. Guhr, A. M\"uller-Groeling and H.A. Weidenm\"uller,
cond-mat/9707301, Phys. Rep. (in press).

\bibitem{Beenreview}C. Beenakker, Rev. Mod. Phys. {\bf 69}, 731 (1997).

\bibitem{Montambaux}G. Montambaux, cond-mat/9602071.

\bibitem{Altshuler} B. Altshuler, I. Zharekeshev, S. Kotochigova and
B. Shklovskii, Zh. Eksp. Teor. Fiz. {\bf 94}, 343 (1988).

\bibitem{Altland} A. Altland and Y. Gefen, Phys. Rev. Lett. {\bf 71}, 3339 
(1993); D. Braun and G. Montambaux, Phys. Rev. {\bf 52}, 13903 (1995);
A. Aronov and A. Mirlin, 
Phys. Rev. {\bf B51}, 6131 (1995);
Y. Fyodorov and A. Mirlin, Phys. Rev. {\bf B51}, 13403 (1995);
T. Guhr and A. Mueller-Groeling,  cond-mat/9702113;
V. Kravtsov, I. Lerner, B. Altshuler and A. 
Aronov, Phys. Rev. Lett. {\bf 72}, 888 (1994);
A. Altland, Y. Gefen and G. Montambaux, 
Phys. Rev. lett. {\bf 76}, 1130 (1996);
N. Argaman, Y. Imry and U. Smilansky, 
Phys. Rev. {\bf B47}, 4440 (1993);
K. Frahm, T. Guhr and A. Mueller-Groeling, cond-mat/9801298.
\bibitem{Chalker-kravtsov}J. Chalker, V. Kravtsov, I. Lerner,
JETP Lett. 64, 386 (1996);
V. Kravtsov and I. Yurkevich, cond-mat/9612036.
\bibitem{BC} T. Banks and A. Casher, Nucl. Phys. {\bf B169}, 103 (1980).
\bibitem{SV}E. Shuryak and J. Verbaarschot,
Nucl. Phys. {\bf A560}, 306 (1993).
\bibitem{Tilo} M. Berbenni-Bitsch, S. Meyer, A. Sch\"afer,
J. Verbaarschot and  T. Wettig, Phys. Rev. Lett. {\bf 80}, 1146 (1998);
R. Pullirsch, K. Rabitsch, T. Wettig and H. Markum, hep-ph/9803285.
\bibitem{Ma}J.Z. Ma, T. Guhr and T. Wettig, hep-lat/9712026.
\bibitem{Vinst}J. Verbaarschot, Nucl. Phys. {\bf B427}, 434 (1994).
\bibitem{Halasz}M. Halasz and J. Verbaarschot,
Phys. Rev. Lett. {\bf 74}, 3920 (1995).
\bibitem{shuryak}E. Shuryak, Nucl. Phys. {\bf B302}, 599 (1988); 
D. Diakonov, hep-ph/9602375; A.V. Smilga, Phys. Rep. 291, 1 (1997).
\bibitem{V} J. Verbaarschot, Phys. Rev. Lett. {\bf 72}, 2531 (1994); 
Phys. Lett. {\bf B329} (1994) 351; Nucl. Phys. {\bf B426[FS]}, 559 (1994).
\bibitem{GL}J. Gasser and H.~Leutwyler, Phys. Lett. {\bf 188B}, 447 (1987).
\bibitem{LS}H.~Leutwyler and A.~Smilga, Phys. Rev. {\bf D46}, 5607 (1992).
\bibitem{vPLB}J. Verbaarschot, Phys. Lett. {\bf B368}, 137 (1996).
\bibitem{Stern}J. Stern, hep-ph/9801282.
\bibitem{osborn}J. Osborn and J. Verbaarschot, hep-ph/9803419.
\bibitem{VZ}J. Verbaarschot and I. Zahed, Phys. Rev. Lett. {\bf 70},
3852 (1993).
\bibitem{Damgaard}G. Akemann,
P. Damgaard, U. Magnea and S. Nishigaki, Nucl. Phys. {\bf B 487[FS]}, 721 
(1997). 
\bibitem{Brezin-Zee}E. Br\'ezin, S. Hikami and A. Zee,
Nucl. Phys. {\bf B464}, 411 (1996).
\bibitem{Dampart}P. Damgaard, hep-th/9711110,  
hep-th/9711047; G. Akemann and P. Damgaard, hep-th/9801133, hep-th/9802174; 
S. Nishigaki, P. Damgaard and T. Wettig, hep-th/9803007;
\bibitem{Tilo-Guhr}T. Guhr and T. Wettig,
Nucl. Phys. {\bf B506}, 589 (1997).
\bibitem{Seneru}A. Jackson, M. Sener and J. Verbaarschot, Nucl. Phys.
{\bf B479}, 707 (1996).
\bibitem{SS97}T. Sch\"afer and E. Shuryak,  Rev. Mod. Phys. {\bf 70}, 323
(1998).
\bibitem{Guhrp}T. Guhr, private communication.
\bibitem{Janssen}K. Jansen and C. Liu, Nucl. Phys. Proc. Suppl. {\bf 53}, 974
(1997).
\bibitem{Toublan}J. Osborn, D. Toublan and J. Verbaarschot, hep-th/9807110.




\end{references}
\end{document}